\pdfoutput=1
\documentclass[floatfix,altaffilletter,preprintnumbers,superscriptaddress,tightenlines,showpacs,showkeys,nofootinbib]{revtex4-1}

\usepackage[dvips]{graphicx}
\usepackage{array,amsmath,amsthm,feynmf,mathtools}
\usepackage{amssymb}
\usepackage{braket}
\usepackage{breakurl}
\usepackage{color}
\usepackage{enumitem}
\usepackage{epsfig}
\usepackage{graphicx}
\usepackage{hyperref}
\usepackage{mathtools}
\usepackage{siunitx}
\usepackage{slashed}
\usepackage{xcolor}
\usepackage[normalem]{ulem}

\usepackage{lmodern}
\usepackage{tikz}

\usepackage[T1]{fontenc}

\makeatletter
\def\@fpheader{\relax}
\makeatother

\DeclareSymbolFont{AMSa}{U}{msa}{m}{n}
\DeclareSymbolFont{AMSb}{U}{msb}{m}{n}
\DeclareMathSymbol{\fieldR}{\mathalpha}{AMSb}{"52}

\newcommand{\beq}{\begin{eqnarray}}
\newcommand{\eeq}{\end{eqnarray}}
\newcommand{\bea}{\begin{eqnarray}}
\newcommand{\eea}{\end{eqnarray}}
\newcommand{\be}{\begin{equation}}
\newcommand{\ee}{\end{equation}}
\newcommand{\bq}{\begin{equation}}
\newcommand{\eq}{\end{equation}}

\def\6{\partial}

\def\6{\partial}

\usepackage{color}
    \definecolor{darkgreen}{rgb}{0,0.5,0}
    \definecolor{darkblue}{rgb}{0,0,0.6}
    \definecolor{purple}{rgb}{0.4,.2,0.7}

\newcommand{\fig}[1]{Fig.~\ref{#1}}

\newcommand{\eqn}[1]{(\ref{#1})}
\newcommand{\eqq}[1]{Eq.~(\ref{#1})}

\newcommand{\sac}{\, , \qquad}

\makeatletter

\begin{document}                                                                                                                                                                                                                                                                                                                                                                                                                                                                                                                                                                                                                                                                
\begin{flushright}
    CPHT-RR050.062024
\end{flushright}

\title{Hydrodynamics of Relativistic Superheated Bubbles}

\author{Yago~Bea}
\affiliation{Departament de F\'\i sica Qu\`antica i Astrof\'\i sica (FQA),  Universitat de Barcelona (UB), Mart\'\i\  i Franqu\`es, 1, 08028 Barcelona, Spain.}
\affiliation{Institut de Ci\`encies del Cosmos (ICCUB),  Universitat de Barcelona (UB), Mart\'\i\  i Franqu\`es, 1, 08028 Barcelona, Spain.}
\author{Jorge~Casalderrey-Solana}
\affiliation{Departament de F\'\i sica Qu\`antica i Astrof\'\i sica (FQA),  Universitat de Barcelona (UB), Mart\'\i\  i Franqu\`es, 1, 08028 Barcelona, Spain.}
\affiliation{Institut de Ci\`encies del Cosmos (ICCUB),  Universitat de Barcelona (UB), Mart\'\i\  i Franqu\`es, 1, 08028 Barcelona, Spain.}
\author{David~Mateos}
\affiliation{Departament de F\'\i sica Qu\`antica i Astrof\'\i sica (FQA),  Universitat de Barcelona (UB), Mart\'\i\  i Franqu\`es, 1, 08028 Barcelona, Spain.}
\affiliation{Institut de Ci\`encies del Cosmos (ICCUB),  Universitat de Barcelona (UB), Mart\'\i\  i Franqu\`es, 1, 08028 Barcelona, Spain.}
\affiliation{Instituci\'o Catalana de Recerca i Estudis Avan\c cats (ICREA), Passeig Llu\'\i s Companys 23,  08010 Barcelona, Spain.}
\author{Mikel Sanchez-Garitaonandia}
\affiliation{CPHT, CNRS, \'Ecole polytechnique, Institut Polytechnique de Paris, 91120 Palaiseau, France}

\begin{abstract}
Relativistic, charged, superheated bubbles may play an important role in neutron star mergers if first-order phase transitions are present in the phase diagram of Quantum Chromodynamics. We describe the properties of these bubbles in the hydrodynamic regime. We find two qualitative differences with supercooled bubbles. First, the pressure at the center of an expanding superheated bubble can be higher or lower than the pressure in the asymptotic, metastable phase. Second, some fluid flows develop  metastable regions behind the bubble wall for any choice of the equation of state. We consider the possible role of a conserved charge akin to baryon number. The fluid flow profiles are unaffected by this charge if the speed of sound is constant in each phase, but they are modified for more general equations of state. We compute the efficiency factor relevant for gravitational wave production. 

\end{abstract}

\maketitle
\section{Introduction}
\label{intro}
\begin{figure}[b]
    \begin{center}
    	\includegraphics[width=0.95\textwidth]{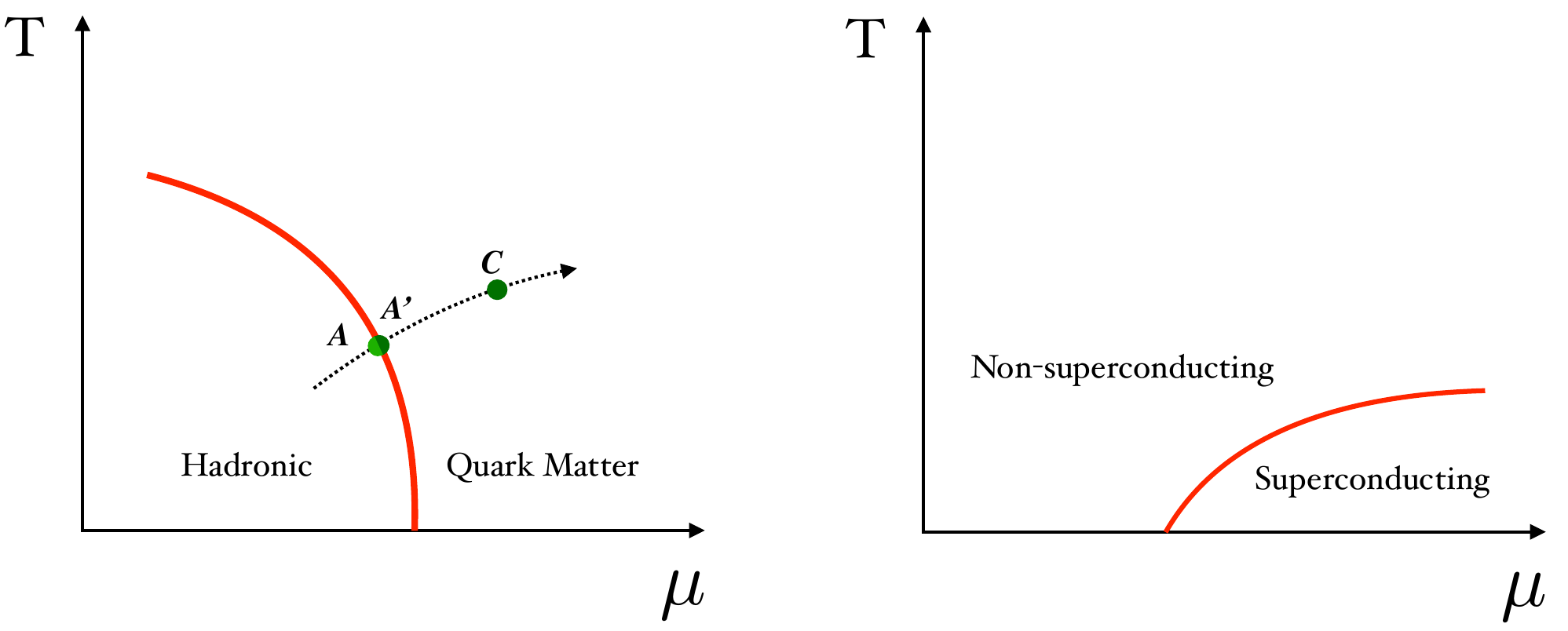}
    \end{center}
    \caption{\small Two possible phase transitions in QCD, indicated by the solid red curves.  $T$ and $\mu$ are the temperature and the baryon chemical potential, respectively.  The dotted black curve on the left panel shows a possible evolution of a region of a NS merger as this region is heated and/or compressed. 
    The points dubbed $A$, $A'$ and $C$ correspond to the states shown in \fig{meta}. See text and Ref.~\cite{Casalderrey-Solana:2022rrn}.}
\label{FOPTquark}
\end{figure}

A  variety of arguments suggest that at least two first-order phase transitions (FOPT) may be present in the phase diagram of Quantum Chromodynamics (QCD) as a function of temperature and baryon chemical potential  \cite{Stephanov:2004wx,Alford:2007xm,Fukushima:2010bq,Guenther:2022wcr}, as sketched in \fig{FOPTquark}. One is the transition from hadronic matter to quark matter. The other is the transition from a non-superconducting to a color-superconducting phase. While these transitions are well motivated, rigorously establishing their existence is a fundamental open problem in nuclear and particle physics whose solution has resisted both theoretical and experimental attempts for decades.

Gravitational waves produced in neutron star (NS) mergers could  provide direct experimental access to these phase transitions \cite{Casalderrey-Solana:2022rrn}. Numerical simulations of NS mergers based on equations of state (EoS) with a hadronic-quark matter phase transition include \cite{Most:2018eaw,Most:2019onn,Ecker:2019xrw,Prakash:2021wpz,Weih:2019xvw,Tootle:2022pvd}. These studies have shown that the dynamics of the merger results in the formation of regions in which the matter is sufficiently heated and/or compressed  that the thermodynamically preferred phase is the quark-matter phase. In other words, the matter in these regions is pushed along the black dotted curve in \fig{FOPTquark}. For brevity, we will refer to these regions simply as ``superheated regions''. 
Although no simulation based on an EoS with a color-superconducting phase has been performed, the large baryon densities found in existing simulations make it conceivable that color-superconducting matter may also be formed in NS mergers.

\begin{figure}[tbp]
\begin{center}
    \includegraphics[width=0.75\textwidth]{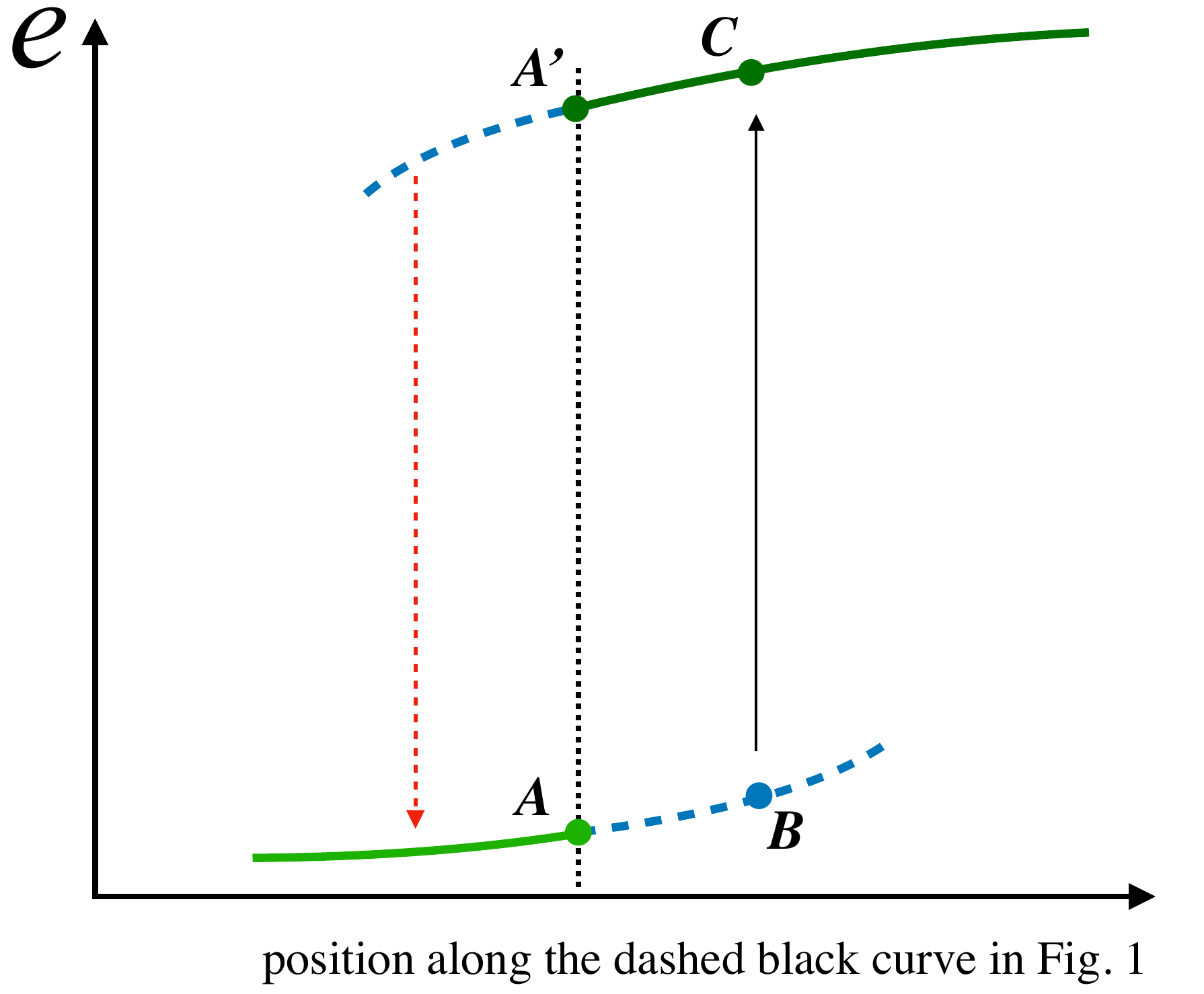}
  \caption{\small Energy density as a function of the position along the  black dotted curve in \fig{FOPTquark}. Both $T$ and $\mu$ increase from left to right. The black, dotted,  vertical line indicates the location of the phase transition, 
   determined by the condition that the states $A$ and $A'$ have the same free energy density. The green and blue curves indicate stable and metastable states, respectively. As some region of the NS merger is sufficiently  heated and/or compressed,  it enters the lower metastable branch. At the point $B$ bubbles of the preferred state $C$ on the upper stable branch are  nucleated. The direction of this phase transition is the opposite of that in a supercooled phase transition, which would take place as indicated by the vertical, dotted, red arrow.
   }
   \label{meta}
\end{center}
\end{figure}

The energy density along the black dotted curve in \fig{FOPTquark} displays the multivalued form characteristic of a FOPT, as  shown in \fig{meta}. The superheated region is pushed from left to right along the lower branch of the phase diagram. Once the superheating is large enough, namely once the region of interest is sufficiently deep into the metastable branch, bubbles of the stable phase begin to nucleate.\footnote{To avoid confusion, we clarify that by ``bubble'' we mean the \emph{volume} of a given phase separated from the outside phase by the ``bubble wall'', which is a \emph{surface}.} The point where this happens is labeled ``$B$'' in  \fig{meta}, the nucleated phase is labelled ``$C$'', and the transition is indicated by a vertical, solid, black arrow. We will refer to these nucleated bubbles as ``superheated bubbles''. The dynamics of these bubbles can produce gravitational waves that would provide direct experimental access to the QCD phase transition \cite{Casalderrey-Solana:2022rrn}. The order-of-magnitude estimate in this reference shows that the frequency of these gravitational waves falls roughly in the MHz range, and that they may be potentially observable with future detectors. 

The direction of the ``superheated'' transition is opposite to that of a cosmological FOPT. In this case the Universe is supercooled, namely pushed from right to left along the upper branch into the upper metastable region, until it transitions down to the lower stable branch, as indicated in \fig{meta} by the dashed, vertical, red arrow. We will refer to the nucleated bubbles in this case as ``supercooled bubbles''. This type of transition will also occur in a NS merger if the superheated region cools down again.

The dynamics of relativistic supercooled bubbles has been extensively studied in the literature, motivated  by their potential implications for cosmological FOPT
--- see e.g.~\cite{Hindmarsh:2020hop} for a review.  Motivated by their possible role in NS mergers, in this paper we initiate the study of relativistic, charged, superheated bubbles, focusing on their hydrodynamic properties in the simple case of a bag-model, conformal EoS. We find two qualitative differences compared to supercooled bubbles. First, the pressure inside a superheated bubble may be higher or lower than the pressure outside the bubble. In contrast, in supercooled bubbles the inside pressure is always higher than the outside pressure.  
By ``inside'' and ``outside'' we do not mean ``right behind'' and ``right in front'' of the bubble wall, but deep inside the bubble (in the stable phase) and asymptotically far away from the bubble (in the metastable phase) --- see Fig.~\ref{fig:bubbles_conformal_p_and_e}(top). The fact that hydrodynamics allows this pressure difference to take either sign in superheated bubbles is remarkable because it shows that the usual intuition, according to which the bubble expands because the pressure inside wins over the combination of outside pressure  plus surface tension, is overly simplistic. 
The second difference is that some fluid flows develop regions behind the bubble wall that are necessarily metastable which would therefore decay at sufficiently long times. 
As we will explain, this is not true for supercooled transitions.

The physics of relativistic and non-relativistic bubbles is expected to be qualitatively similar if the two phases that determine the phase transition have comparable compressibilities in the range of pressures accessed by the bubble dynamics. 
For example, relativistic supercooled  bubbles find a non-relativistic analog in non-relativistic chemical combustions, for this reason often referred to as ``relativistic combustions''. Thus, one may expect that these similarities are also present for  relativisitic superheated bubbles. However, a qualitative difference appears in a particular case of obvious phenomenological interest: the nucleation and expansion of vapor bubbles in a superheated liquid. In this case the liquid is effectively incompressible, and this leads to a bubble wall velocity that decreases as $t^{-1/2}$ at asymptotically late times. For completeness, we review this case in Appendix \ref{AppA}. 

Part of the contents of this paper overlap with those in Ref.~\cite{Barni:2024lkj}, which appeared while this paper was being typewritten. To facilitate comparison with that reference, we note that what we call deflagrations, hybrids and detonations are dubbed inverse detonations, inverse hybrids and inverse deflagrations in \cite{Barni:2024lkj}.

\section{Self-similar profiles}

At late times, when the bubbles have become macroscopic in size, the lack of any macroscopic scale implies that these bubbles are expected to become self-similar. This means that the fluid flow profile is a  function of only $\xi = r/t$, with $r$ the radius of the spherical bubble. Besides the bubble wall and possible shocks in the system, the very diluted gradients mean that a 
perfect-fluid description should be a valid approximation of the plasma. In this case, the constitutive equations for the stress tensor and the conserved charge take the form
\begin{equation}
T^{\mu\nu} = wu^{\mu}u^{\nu}+p g^{\mu\nu}, \quad\quad j^{\mu}  = n u^{\mu}\,,
\label{eq:Conservation_laws}
\end{equation}
with $p$  the pressure, $w=e+p$ the enthalpy, $e$ the energy density, $n$ the charge density of the conserved charge and $u^{\mu} = \gamma (1,\vec v)$ the fluid 
four-velocity. The thermodynamic variables are not all independent but are related to the the temperature $T$ and chemical potential of the plasma $\mu$ through the EoS. In this section we will not assume any specific EoS (but later in this section we will assume a constant speed of sound). 

The equations of motion are simply the conservation equations of the stress-tensor and the current:
\begin{equation}
\partial_{\mu}T^{\mu\nu} = 0\,, \quad \quad \partial_{\mu} j^{\mu} = 0\,.
\end{equation}
As in the neutral case (see e.g.~\cite{Espinosa:2010hh}), we can project the motion of the flow along the direction parallel and orthogonal to the velocity of the fluid. For this purpose  we multiply the conservation equations by $u^{\mu} = \gamma(1,\vec{v})$ and $\bar{u}^{\mu}=\gamma(v,\vec{v}/v)$. Using that $u_{\mu}\partial_{\nu}u^{\mu}=0$, we obtain the following equations:
\begin{equation}
\begin{aligned}
w\, \partial_{\mu}u^{\mu}+ u^{\mu}\,  \partial_{\mu}e = 0 &\,, \\[2mm]
w\, \bar{u}^{\nu}u^{\mu} \, \partial_{\mu}u_{\nu}-\bar{u}^{\mu}\, \partial_{\mu}p = 0 &\,,\\[2mm]
n\, \partial_{\mu}u^{\mu} + u^{\mu}\, \partial_{\mu}n 
= 0 &\,.
\end{aligned}
\end{equation}
The self-similarity assumption implies that 
\begin{equation}
\begin{aligned}
& u^{\mu}\partial_{\mu} = -\frac{\gamma}{t}(\xi-v)\partial_{\xi}\,,\\[2mm]
& \bar{u}^{\mu}\partial_{\mu} = \frac{\gamma}{t}(1-\xi v)\partial_{\xi}, \\[2mm] 
& \partial_{\mu}u^{\mu} = (1-\xi v)\frac{\gamma^3}{t}\partial_{\xi} + \frac{\gamma v}{r}(d-1),
\end{aligned}
\end{equation}
where $d$ is the number of spatial dimensions. Substituting these expressions into the equations of motion leads to the following set of differential equations:
\begin{eqnarray}
   (\xi-v)\frac{\partial_{\xi}e}{w} &=& \gamma^2(1-\xi v)\partial_{\xi}v+(d-1)\frac{v}{\xi},\\[2mm]
(1-\xi v)\frac{\partial_{\xi}p}{w} &=& \gamma^2(\xi-v)\partial_{\xi}v,\\[2mm]
(\xi-v)\frac{\partial_{\xi}n}{n} &=& \gamma^2(1-\xi v)\partial_{\xi}v+(d-1)\frac{v}{\xi}\,. 
\end{eqnarray}
These can be recast into a more convenient form:
\begin{subequations}
    \begin{align}
    \label{v_eq}
&\gamma^2(1-\xi v)\left(\frac{\nu(\xi,v)^2}{c_s^2}-1\right)\partial_{\xi}v = (d-1)\frac{v}{\xi}\,,\\[2mm]
\label{w_eq}
&\frac{\partial_{\xi}w}{w} = \gamma^2\left(1+\frac{1}{c_s^2}\right)\nu(\xi,v)\partial_{\xi}v\,, \\[2mm]
\label{n_eq}
&\frac{\partial_{\xi}n}{n}  = \frac{1}{1+c_s^2}\frac{\partial_{\xi}w}{w}\,,
    \end{align}
    \label{eq:Self_Similar_Flow}
\end{subequations}
where 
\begin{equation}
\label{cs}
    c_s^2 = \partial_{e}p+\frac{n}{w}\,  \partial_np
\end{equation}
is the speed of sound and 
\begin{equation}
  \nu(\xi,v)=\frac{\xi-v}{1-\xi v}  
\end{equation}
is the Lorentz-boosted velocity. 

In order to obtain a closed system of equations, we must specify a relation between pressure, energy density and charge density, $p(e, n)$. This is sometimes referred to as the ``reduced EoS'' (see e.g.~\cite{Komoltsev:2021jzg}) because, although it determines the hydrodynamic flows, we will see in Sec.~\ref{sols} that it contains less information than the true EoS, $p(T,\mu)$. If the pressure in each phase obeys $p=e/d$, then the speed of sound is constant, $c_s^2=1/d$. In this case, \eqref{eq:Self_Similar_Flow} becomes a nested system that can be sequentially solved for $v$, $w$ and $n$, similarly to what happens in the neutral case \cite{Espinosa:2010hh}. The nested structure means that the addition of charge does not change the properties of self-similar flows for conformal EoS. In fact, the last equation simply implies that $n\propto w^{1/(1+c_s^2)}$. 
In contrast, if the speed of sound is not constant, $c_s=c_s(e,n)$, then equations \eqref{eq:Self_Similar_Flow} are coupled to one another. In this case the effect of the charge is already present at the level of the self-similar flow. We leave studies of non-conformal theories for future work.

At the location of the bubble wall or of the possible shocks that will form during the expansion of the bubble, the perfect fluid description will break down. However, we can integrate the conservation laws, in the radial direction, in the rest frame of these discontinuities. Doing so leads to the following matching conditions: 
\begin{equation}
\begin{aligned}
w_+ v_+^2 \gamma_+^2+p_+ &= w_- v_-^2 \gamma_-^2+p_- \,,\\[2mm]
w_+ v_+ \gamma_+^2 &= w_- v_- \gamma_-^2 \,,\\[2mm]
n_+ v_+ \gamma_+ & = n_- v_- \gamma_- \,,
\end{aligned}
\label{eq:matchingc}
\end{equation}
where $+$($-$) refers to the plasma in front (behind) the discontinuity, the velocities $v_\pm$ are measured in the rest frame of the discontinuity, and $\gamma$ is the usual Lorentz factor.  Rearranging the first two matching conditions we arrive at 
\begin{equation}
	v_+v_- = \frac{p_+-p_-}{e_+-e_-} \,, 
 \quad\quad 
	\frac{v_+}{v_-} = \frac{e_-+p_+}{e_++p_-}\, , \quad\quad	
	n_+ v_+ \gamma_+ = n_- v_- \gamma_-.
\label{eq:Matching}
\end{equation}
Eqs.~\eqref{eq:Self_Similar_Flow} and \eqref{eq:Matching} determine the entire fluid  flow as a function of three input parameters: the wall velocity $\xi_w$, the nucleation energy density  $e_n$ and the nucleation charge density $n_n$. 

In the following we will solve these equations in the case of spherical bubbles in three spatial dimensions ($d=3$) and a constant speed of sound. 

\begin{figure}
	\centering
	\includegraphics[width=0.85\textwidth]{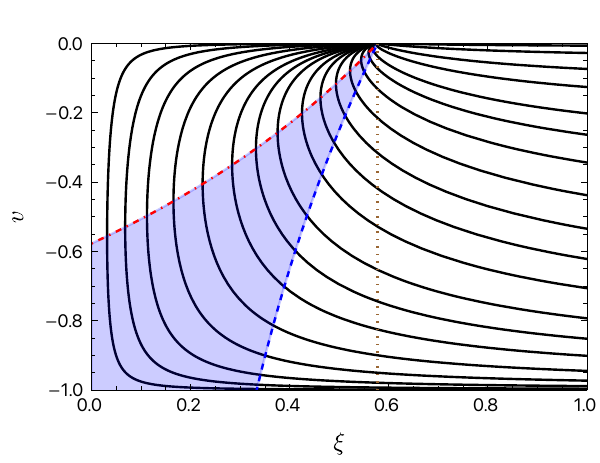}
	\caption{Self-similar flows for superheated bubbles in the case $c_s^2=1/3$. The vertical dotted line lies at $\xi=c_s$, marking the beginning of detonations. The dashed blue curve corresponds to the locus  $\nu(\xi,v)\xi=c_s^2$, where the shock must be located . The dotted-dashed red curve indicates the points where the boosted velocity $\nu(\xi,v)=c_s$ and $\partial_{\xi}v\rightarrow\infty$. Expanding bubbles in the shaded area are not allowed.}
	\label{fig:all_self_similar_flows}
\end{figure}

When $c_s$ is constant we can solve for the velocity in \eqref{eq:Self_Similar_Flow} independently of the rest of the variables. By assuming different initial conditions $v(\xi_0)=v_0$, one can obtain all possible flows that are self-similar, shown in Fig.~\ref{fig:all_self_similar_flows}. Contrary to the supercooled case, we seek  solutions where at $\xi>0$ (expanding bubbles) the flow has negative speed $v<0$ (energy inflow). In order to produce Fig.~\ref{fig:all_self_similar_flows}, we chose $c_s^2=1/3$. We expect that different values of the speed of sound will  change  the picture quantitatively but not qualitatively.

For $v\rightarrow 0$, all flows tend to $\xi=c_s$, doing so from low to high values of $\xi$. This implies that flows that start at $\xi<c_s$ can continuously connect to a $v=0$ state, while flows starting at $\xi>c_s$ will inevitably have to develop a discontinuity to do so. Moreover, in the region $\xi<c_s$
another discontinuity may occur, 
for which the matching conditions reduce to $\nu(\xi,v)\xi=c_s^2$. This discontinuity is a shock and its location is displayed in Fig.~\ref{fig:all_self_similar_flows} as a dashed blue curve.

With this information  we  can already see the type of fluid flows that we can get. For low velocity we expect a so-called ``deflagration'', namely a flow in front of the wall that smoothly connects with the asymptotic state at rest. For high velocity, we expect a ``detonation'', namely a flow behind the wall that will develop a shock, connecting with the state at rest inside the bubble. Finally, we can also have
hybrid solutions, which combine features from both deflagrations and detonations. Hybrid flows consist of a flow behind the wall (as in a detonation) ending in a shock, plus a flow ahead of the wall that smoothly connects with $v=0$ (as in a deflagration). 

There is yet another kind of flows, those that are contained in the shaded area of Fig.~\ref{fig:all_self_similar_flows}. Given that they would become multivalued, these flows cannot cross the dot-dashed line, and they must follow to higher $\xi$. As they cannot smoothly connect to a state with $v=0$, they jump discontinuously to it at the shock location. This shock is different to the one present in hybrids and detonations as the flow in motion is behind and not ahead. An entropic analysis shows that the shock front in Fig.~\ref{fig:all_self_similar_flows} increases entropy only if the flow in motion is ahead. Hence, no flow starting in the mentioned area is physically realizable.

In the next subsection we explore in more detail the three type of solutions just mentioned.

\section{Solutions for a conformal equation of state}
\label{sols}

Hereafter we focus on the case of three spatial dimensions, $d=3$. In order to analyze the  fluid flows in more detail, a choice of an EoS is needed.

Assuming two branches such that $c_s^2=1/3$ in each of them, we choose
\begin{equation}
\begin{aligned}
		p_H(T,\mu) & = a_H T^4 +b_H T^2 \mu^2 +c_H \mu^4-\epsilon\,, \\[2mm]
		p_L(T,\mu) & = a_L T^4 +b_L T^2 \mu^2 +c_L \mu^4\,,
		\label{eq:pressures}
\end{aligned}
\end{equation}
where $a_{H,L}$, $b_{H,L}$ and $c_{H,L}$ reflect the number of degrees of freedom in each phase and ``$H$'' and ``$L$'' refer to the high- and the low-energy phases, respectively. The constant $\epsilon>0$ is sometimes referred to as the ``bag constant'' and it measures the energy difference between the two phases in vacuum, that is, at $T=\mu=0$. 
This is easily verified by means of the usual thermodynamic relations
\be
e + p = Ts +\mu n \sac
s= \frac{\partial P}{\partial T}\sac 
n= \frac{\partial P}{\partial \mu}\,,
\ee
which for the two branches in \eqn{eq:pressures} yield
\be
e_H = 3 p_H + \epsilon \sac 
e_L = 3 p_L \,.
\label{reduced}
\ee
Eqs.~\eqn{reduced} are what we referred to above as the ``reduced EoS''. Note that different microscopic theories give rise to different EoS through different values of the constants $a_{H,L}$, $b_{H,L}$ and $c_{H,L}$, but  all EoS of the form \eqn{eq:pressures} lead to the same reduced EoS \eqn{reduced}. 

The pressure or the energy density are defined up to an overall, unphysical, additive constant $p_0$. In other words, in the absence of gravity the physics is invariant under the simultaneous shift\footnote{In the presence of gravity, e.g.~in Cosmology, $p_0$ becomes physical since it affects the expansion rate of the Universe.} 
\be
p_{H} \to p_{H}+p_0 \sac  
p_{L} \to p_{L}+p_0 \,.
\ee
Here we have fixed this ambiguity without loss of generality by setting 
\be
p_0=0 \,.
\label{p0}
\ee
This is equivalent  to requiring that the pressure vanish in the vacuum state, namely that
\be
p_L(T=0,\mu=0)=0 \,.
\ee
With this convention, the global stability of the vacuum state constraints the parameters $a_{H,L}$, $b_{H,L}$ and $c_{H,L}$ that specify the EoS so that 
\be
p_L\geq 0 \,. 
\label{pos}
\ee

In contrast to $p_0$, the bag constant is physical because it measures the pressure or energy difference between the two phases. This introduces a preferred scale that breaks conformal invariance, as reflected for example by the fact the trace of stress tensor, $e-3p$, is non-zero in the $H$-phase. This scale sets the overall magnitude of the critical temperature and chemical potential at which the phase transition between the two branches takes place, which is determined by the condition 
\be
\label{phasetrans}
p_L(T_c,\mu_c) = p_H(T_c,\mu_c) \,.
\ee

We thus see that referring to \eqn{eq:pressures} as a conformal equation of state, as is common in the literature, is an abuse of language. The reason for this terminology is the fact that the speed of sound \eqn{cs} is 1/3 in both branches, and that the bag constant cancels out in the enthalpy $w=e+p$. In the hydrodynamic context, since only $c_s$ and $w$ enter the equations \eqn{eq:Self_Similar_Flow}, this means that  the flows on both sides of the bubble wall 
are locally the same as in a conformal theory. Conformality is  broken at the bubble wall, where the  jump implied by the matching conditions \eqn{eq:matchingc}-\eqn{eq:Matching} does depend on the bag constant. The entire flow is thus governed  solely by the reduced EoS. As a consequence, microscopic theories with different values of the constants $a_{H,L}$, $b_{H,L}$ and $c_{H,L}$ give rise to the same set of flows. However, the stability of these flows does depend on these constants, since they determine the location of the phase transition and hence of the boundaries between stable, metastable and unstable regions, as well as determining if entropy increases or not during the expansion. Therefore, a given flow may be stable according to one EoS and unstable according to another. We will come back to this point below. 

\begin{figure}
	\centering
	\includegraphics[scale=1 ]{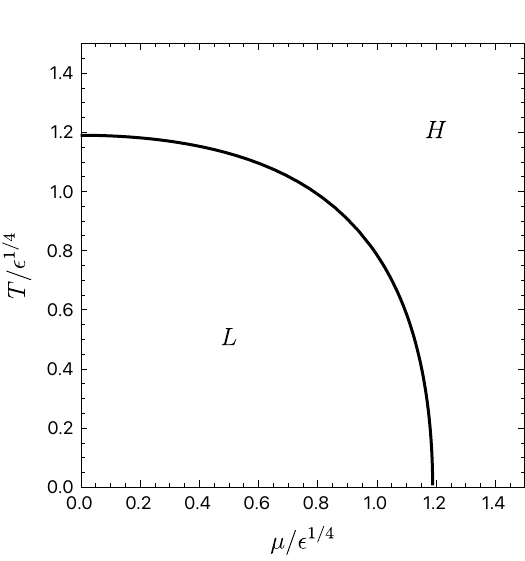}
	\caption{Phase diagram for the choice $a_H=b_H=c_H=1$ and $a_L=b_L=c_L=1/2$.  ``$L$'' and ``$H$'' refer to the low- and high-energy phases,  respectively. All quantities are measured in units of the bag constant, $\epsilon$, which is the energy difference between the two phases at $T=\mu=0$.}
	\label{fig:phase_diagram_conformal}
\end{figure}

In this paper we will choose 
\be
a_H=b_H=c_H=1 \sac a_L=b_L=c_L=1/2\,, 
\ee
for which the phase diagram is shown in Fig.~\ref{fig:phase_diagram_conformal}. There is a curve of FOPT transition at $T_c^2+\mu_c^2=\sqrt{2\epsilon}$. 
Indeed, recall that the thermodynamically preferred phase is the one with the highest pressure (equivalently, the lowest free energy), and we have 
\be
p_L > p_H \,\,\,\,\, \mbox{if} \,\,\,\,\, T_c^2+\mu_c^2 < \sqrt{2\epsilon}\sac p_H > p_L \,\,\,\,\, \mbox{if}\,\,\,\,\, T_c^2+\mu_c^2 > \sqrt{2\epsilon}\,.
\label{pp}
\ee

We do not choose this EoS because we expect it to provide a realistic description of  QCD but because it is the simplest example in which an investigation of superheated bubbles is possible. We will report on more realistic examples elsewhere. 

Having specified an EoS, we can now solve  
Eqs.~\eqref{eq:Self_Similar_Flow} together with the matching conditions \eqref{eq:Matching} in order to obtain the entire flow. 

\begin{figure}
	\centering
 
	{\includegraphics[scale=0.75]{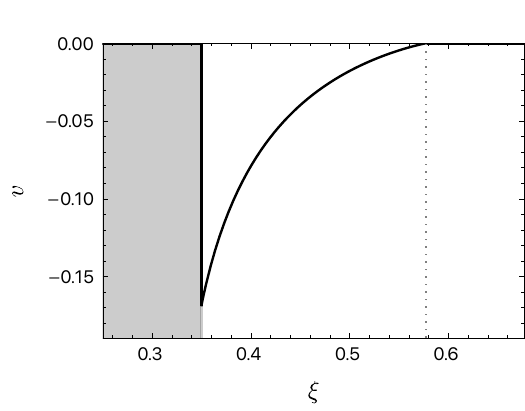}
		\hspace{5mm} \includegraphics[scale=0.75]{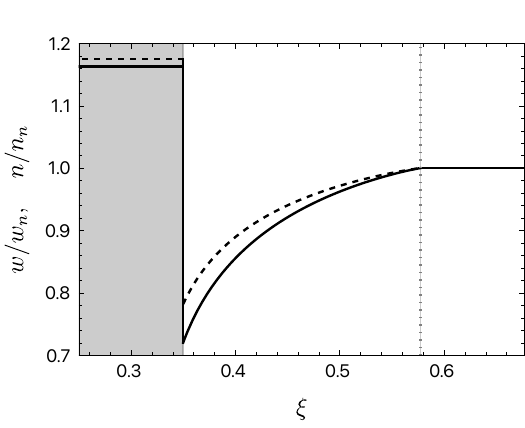}
	}

	{\includegraphics[scale=0.75]{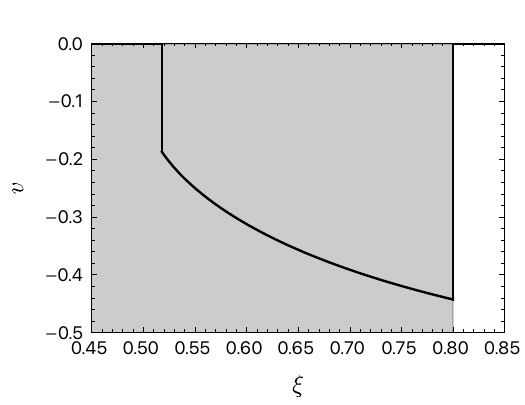}
	\hspace{5mm}  \includegraphics[scale=0.75]{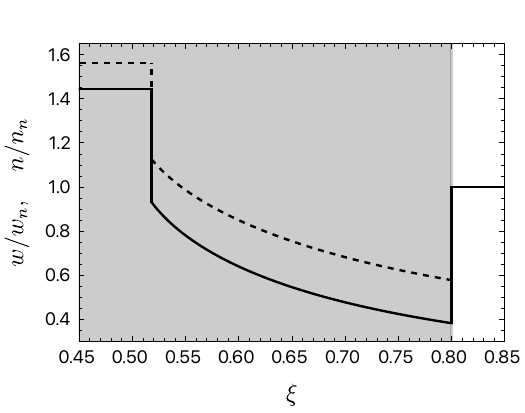}
}

	{\includegraphics[scale=0.75]{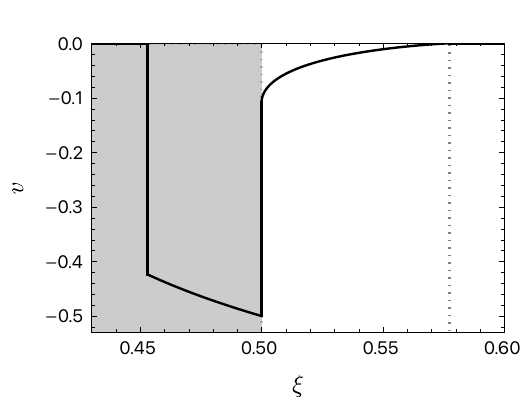}
	\hspace{5mm}  \includegraphics[scale=0.75]{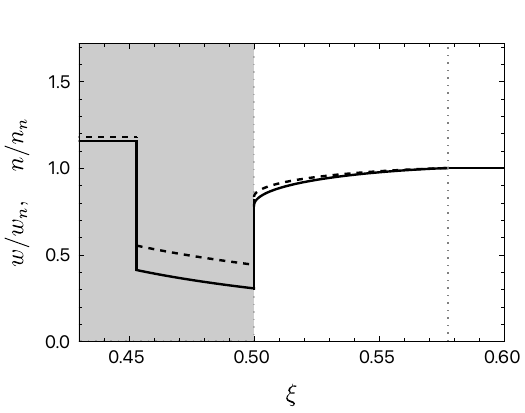}
}
\caption{Fluid flows for superheated bubbles in the case of deflagrations (top), detonations (middle) and hybrids (bottom), with parameters $\xi_w=\{0.35,0.8,0.5\}$ and $\alpha_n=\{0.0614,0.143,0.0421\}$, respectively. The left panel shows the fluid velocity. The right panel shows the enthalpy (solid-black) and the charge density (dashed-black) normalized to their values at infinity. Grey regions correspond to the $H$-phase and white regions to the $L$-phase, with the boundary being the bubble wall. Vertical dotted lines indicate  the speed of sound.
}
\label{fig:bubbles_conformal}
\end{figure}

\begin{figure}
	\centering

	{\includegraphics[scale=0.75]{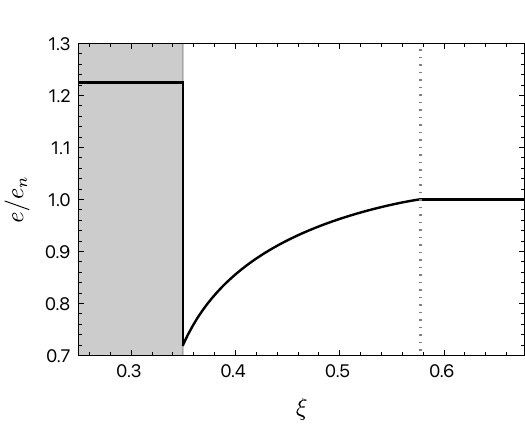}
		\hspace{5mm} \includegraphics[scale=0.75]{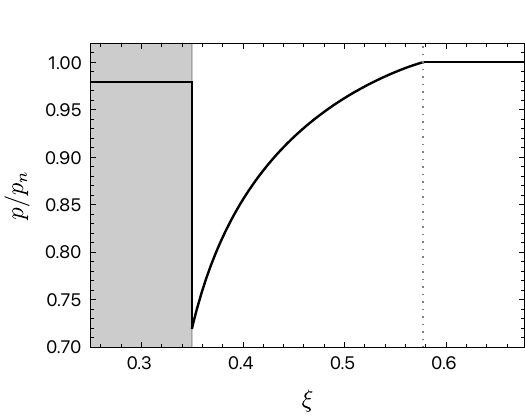}
	}

	{\includegraphics[scale=0.75]{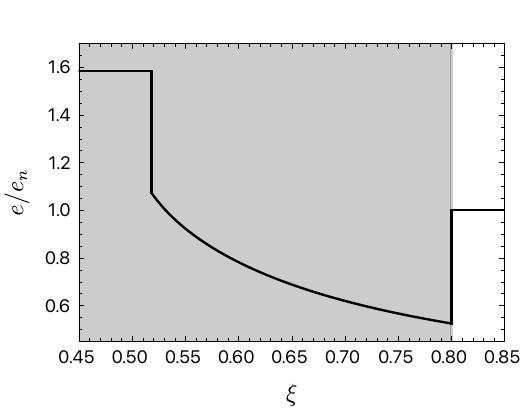}
	\hspace{5mm}  \includegraphics[scale=0.75]{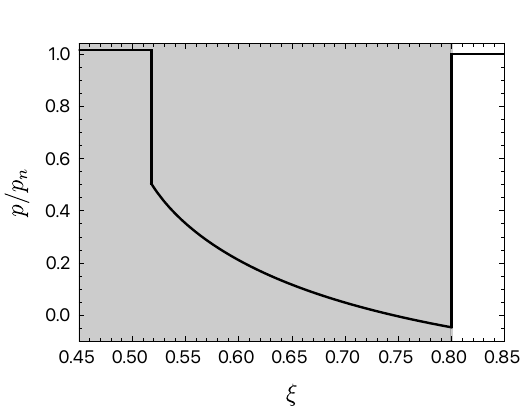}
}

	{\includegraphics[scale=0.75]{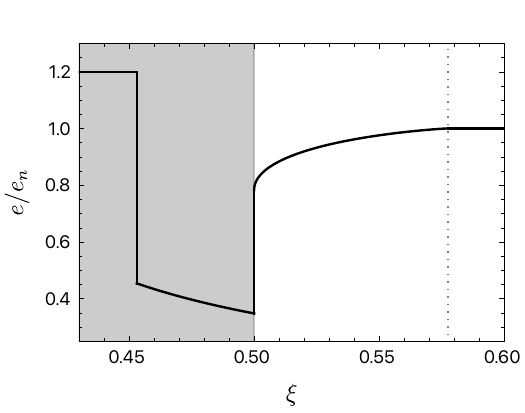}
	\hspace{5mm}  \includegraphics[scale=0.75]{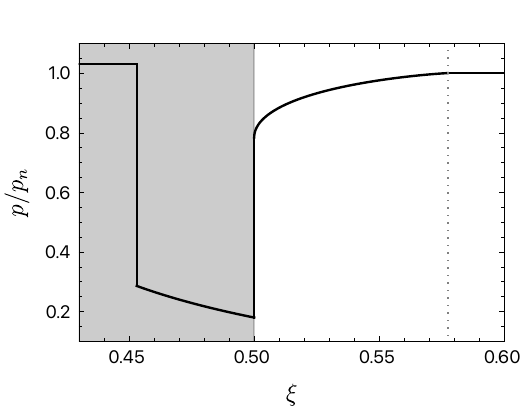}
}
\caption{Energy density (left) and pressure (right) profiles, normalized to their values at infinity, of the fluid flows shown in Fig.~\ref{fig:bubbles_conformal}.}
\label{fig:bubbles_conformal_p_and_e}
\end{figure}

As mentioned earlier, for low, subsonic wall velocities $\xi_w<c_s$ we have deflagration-type solutions. These are characterized by the fact that the fluid behind the wall is at rest. In contrast, the fluid in front of the wall gets into motion perturbed by the movement of the wall. The fluid in motion continuously connects at large distances with the superheated fluid at rest. An example of a deflagration is shown in the top panels of Figs.~\ref{fig:bubbles_conformal} and \ref{fig:bubbles_conformal_p_and_e}. Contrary to what happens in the supercooled case \cite{Espinosa:2010hh}, the flow velocity is negative everywhere and, notably, there is no shock present in the system. The fluid flow in front of the wall continuously connects to the metastable state at $\xi=c_s$. 

In the case of a high, supersonic wall velocity $\xi_w >c_s$, the solution will be of detonation type. In this case, the wall is moving too fast for the fluid ahead to respond and so it remains at rest. As the wall passes the fluid is set in motion and, as it cannot smoothly connect with a state of $v=0$, it develops a shock down the stream. The shock then allows to connect with a fluid at rest inside the bubble. An example of a detonation is depicted in the middle panels of Figs.~\ref{fig:bubbles_conformal} and \ref{fig:bubbles_conformal_p_and_e}. As in the case of deflagrations, the detonation flow differs qualitatively from the supercooled case. In the latter, a rarefaction wave gets formed behind the wall, continuously connecting with the state at rest inside the bubble at $\xi_w=c_s$. Instead, in the current case the fluid dragged by the motion of the wall develops a shock that connects with the state inside the bubble.

Finally, for wall velocities close to but below the speed of sound, $c_s^2 \leq \xi_w \leq c_s$, a new type of flow can be constructed combining both the flows observed in detonations and deflagrations, giving rise to a hybrid solution. An example of this is displayed in the bottom panels of Figs.~\ref{fig:bubbles_conformal} and \ref{fig:bubbles_conformal_p_and_e}. The defining feature is that, in the rest frame of the bubble wall, the flow in front of it moves at the speed of sound $v_+=c_s$. As in any deflagration, the flow ahead of the wall continuously connects with the asymptotic state while the flow behind develops a shock that takes the fluid to rest. Hybrids appear in the current case for subsonic wall speeds, in contrast with  the supercooled case \cite{Espinosa:2010hh}.

The deflagration shown in \fig{fig:bubbles_conformal_p_and_e}(top) is an example of a superheated flow in which the pressure inside the bubble (at $\xi=0$) is lower than the pressure outside the bubble (at $\xi=1$), whereas the detonations (middle panel) and the hybrids (bottom panel) shown in \fig{fig:bubbles_conformal_p_and_e} are examples in which the  inside pressure is higher than the outside pressure. By constructing a large family of expanding superheated bubbles we have observed that all three types of flows can have either ordering of the pressures --- see Fig.~\ref{fig:max_alpha_vs_xi_wall}.

The specific detonation shown in \fig{fig:bubbles_conformal_p_and_e}(middle) has been chosen to illustrate that, for some flows, the pressure in a region just behind the wall, in this case in the region  $\xi\lesssim 0.8$, can be negative. This is necessarily lower than the pressure of the $L$-phase regardless of the values of the parameters $a_{H,L}$, $b_{H,L}$ and $c_{H,L}$ that specify the EoS --- see Eqn.~\eqn{pos}.\footnote{Recall that we fixed the ambiguity in the overall value of the pressure by setting $p_0$ --- see \eqn{p0}. If we had chosen a different value of $p_0$, then we would have found that the pressure right behind the wall in \fig{fig:bubbles_conformal_p_and_e}(middle) is lower than $p_0$, and we would also have concluded that the preferred phase is the $L$-phase.} Since the thermodynamically preferred phase is the one with the highest pressure (equivalently, with the lowest free energy), it follows that right behind the wall the preferred phase is the $L$-phase. This can also be understood in terms of the temperature and the chemical potential. Since $p<0$, the state right behind the wall is in the $H$-phase and $T,\mu$ must obey $T^2 + \mu^2 < \sqrt{2\epsilon}$. Looking at \eqn{pp} we conclude again that, under these conditions,  the preferred phase is indeed the $L$-phase. 

In summary, despite the fact that the flow in \fig{fig:bubbles_conformal_p_and_e}(middle) can be realized in many different microscopic theories, it is unstable in all of them. This is in stark contrast with supercooled flows, for which there is always at least one  choice of the parameters $a_{H,L}$, $b_{H,L}$ and $c_{H,L}$ for which the flow is stable (although it may be unstable for other choices). This is illustrated, for example, by the results in \cite{Caprini:2011uz}. Fig.~4 of this reference shows the comparison between $T$ and $T_c$ right behind the wall as a function of $a_-/a_+$ ($a_L/a_H$ in our terminology). For certain values of this ratio the flow behind the wall is unstable, for others it is stable.

We conclude that, in the superheated case, at arbitrarily late times there is an arbitrarily large, approximately homogeneous region behind the wall that is necessarily thermodynamically disfavoured with respect to the $L$-phase at the same temperature and chemical potential. This region will eventually decay by nucleating bubbles of the $L$-phase if the bubble is allowed to expand  for arbitrarily long times. Whether this happens in practice will depend on the nucleation rates in front and behind the bubble wall, on the bubble wall velocities, etc. Nevertheless, from the viewpoint of the dynamics of a single bubble, flows with negative pressure may regarded as thermodynamically unstable. In \fig{fig:max_alpha_vs_xi_wall} these flows correspond to points above the blue curve labelled as $p_-=0$. Note that deflagrations cannot suffer from this problem since the phase right behind the wall is directly the stable $H$-phase.

We have also constructed detonations with $\xi_{w}<c_s$. In these, the wall connects the outside $v=0$ state with some state below the blue line in Fig.~\ref{fig:all_self_similar_flows}, then the flow extends until the blue line and connects with the inside $v=0$ state through a shock. We checked that the entropy production in these solutions can be positive. These solutions could in principle compete with the deflagrations and hybrids. However, they may be unstable and  decay into the hybrids if perturbed, in analogy to what happens in the supercooled case with deflagrations and hybrids for $\xi_w>c_s$ \cite{Espinosa:2010hh}. We leave this stability analysis for future work.

All the flows above can be labeled by  three independent parameters, that is, $T_n$, $\mu_n$ and $\xi_w$ or, equivalently, $e_n$, $n_n$ and $\xi_w$. In the current case in which $p = c_s^2 e \, +\,  \text{constant}$, one can solve for the enthalpy (or energy density) and the fluid velocity first, and then obtain the density for a given choice of $n_n$. Therefore, for each choice of $(e_n,\xi_w)$ we can extract a whole family of analogous solutions parameterized by $n_n$. As in the case of supercooled bubbles, it is useful to characterize this choice in terms of the so-called ``strength factor'' $\alpha_n$. Its general definition is\footnote{We have inverted the order in the numerator with respect to the supercooled case to ensure that it is  positive.} (see e.g. \cite{Hindmarsh:2020hop,Ares:2020lbt})
\begin{equation}
	\alpha_n \equiv \left. \frac{1}{3}\frac{(e_H-3p_H)-(e_L-3p_L)}{w_L}\right\vert_{(T,\mu)=(T_n,\mu_n)}.
\end{equation}
In the present case this reduces to 
\begin{equation}
	\alpha_n = \frac{\epsilon}{e_L(T_n,\mu_n)} = \frac{\epsilon}{e_n} \,,
\end{equation}
so choosing $\alpha_n$ is equivalent to choosing the nucleation energy density in units of $\epsilon$. However, for general EoS  both the strength factor and the energy density must be  simultaneously specified in order to characterize a bubble solution.

The solutions presented in Figs.~\ref{fig:bubbles_conformal} and \ref{fig:bubbles_conformal_p_and_e} depend on the values of $\xi_w$ and $\alpha_n$, but are independent on the values of the constants $\{a_H,b_H,c_H,a_L,b_L,c_L \}$ in \eqref{eq:pressures}. However, the entropy production depends on these constants in \eqref{eq:pressures}, and the solutions will produce positive or negative entropy depending on the values of these constants. Below we comment further on this point.   

\section{Bounds}
\label{bounds}
We will now see that the presence of a forbidden region in \fig{fig:all_self_similar_flows}, together with the existence of a lower bound on the energy density in each branch, imposes bounds on the possible values of the parameters $(\xi_w, \alpha_n)$ that characterize each possible fluid flow.

Let us start with detonations. Given a wall velocity $\xi_w$, consider increasing the strength $\alpha_n$ or, equivalently, decreasing the nucleation energy $e_n$ relative to the bag constant $\epsilon$. This decreases the energy behind the wall $e_-$ and increases the speed $v_-$. 

\begin{figure}
	\centering
\includegraphics[width=0.99\textwidth]{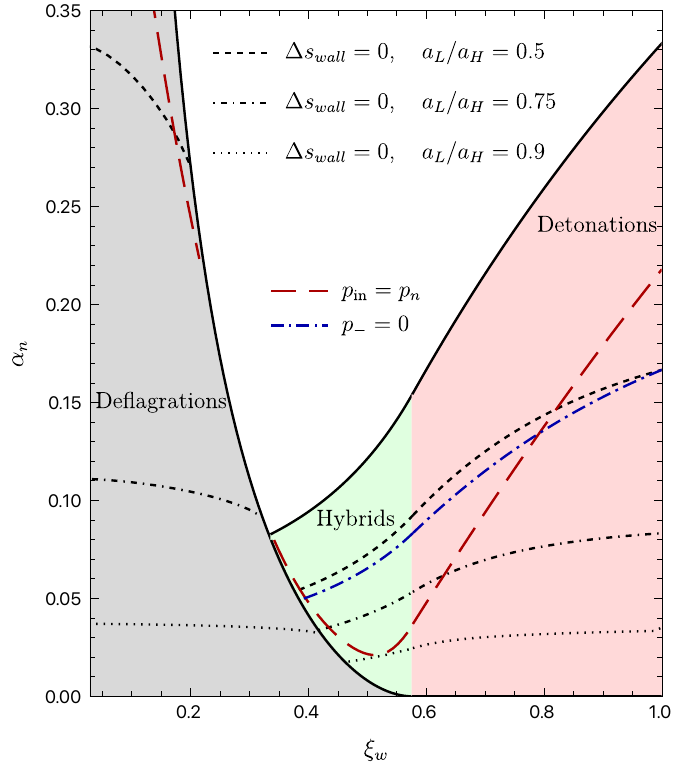}
\caption{Possible fluid flows for different strength factors $\alpha_n$ and wall velocities $\xi_w$. The bounds \eqref{eq:max_strength_detonation}, \eqref{Bound_deflagrations} and \eqref{bound_hibrids} are represented as black solid curves.  The flows allowed by the entropic bound \eqref{Entropy_across_discontinuity} at zero chemical potential for different values of \mbox{$a_L/a_H$} are those below the corresponding discontinuous black curves. The flows for which the pressure inside the bubble is lower (higher) than the pressure outside the bubble are those below (above) the  red curve labelled as $p_{\textrm{in}}=p_n$. The flows for which the fluid pressure becomes negative at some point behind the wall are those above the blue curve labelled as $p_-=0$.}
	\label{fig:max_alpha_vs_xi_wall}
\end{figure}

Clearly, the strongest transition we can consider is that for which the energy density in the fluid behind the wall reaches its vacuum energy, that is, 
$e_- = \epsilon$. Precisely at this point we also reach the maximum possible velocity behind the wall, $v_-=1$. Solving the matching conditions with these values results in the  bound
\begin{equation}
\alpha_{n,\text{det}} \leq \frac{\xi_w-c_s^2}{1+\xi_w}.
\label{eq:max_strength_detonation}
\end{equation}
Detonations are therefore confined to the region simultaneously allowed by the conditions \eqref{eq:max_strength_detonation} and $c_s \leq \xi_w \leq 1$. This region is displayed in pink in Fig.~\ref{fig:max_alpha_vs_xi_wall}.

In the case of deflagrations, a similar upper bound can be obtained for the strength, although it cannot be expressed in terms of a simple analytical formula. Fig.~\ref{fig:all_self_similar_flows} shows that, given a $\xi_w$, the speed ahead of the wall can at most reach the boundary of the excluded region. This happens when the speed of the fluid in front of the wall moves at the speed of sound with respect to the wall, $v_+=c_s$. Notice that this is precisely the value at which hybrid solutions begin to exist. The matching conditions at the wall then lead to the following bound for the state inside the bubble:
\begin{equation}
\alpha_- \leq \frac{(c_s-\xi_w)^2}{c_s^2-2c_s\xi_w+1}.
\label{Bound_deflagrations}
\end{equation}

Translating this bound to one on $\alpha_n$ can only be done numerically, and it requires solving for the entire  fluid flow for deflagrations. The result can be seen in Fig.~\ref{fig:max_alpha_vs_xi_wall}, where deflagrations are confined to the gray shaded region.

Finally, we can obtain a bound on hybrid solutions. Once again, we  see in Fig.~\ref{fig:all_self_similar_flows} that, given a $\xi_w$, the flow behind the wall has to fall below the excluded region. This puts the following constraint on the velocity behind the wall:
\begin{equation}
-1 \leq v(\xi_w) \leq -\frac{c_s^2-\xi_w}{\xi_w(1-c_s^2)}.
\label{bound_hibrids}
\end{equation}
The upper bound, which 
depends on the speed of sound, means that the wall must move faster than the shock, i.e. the flow must lie below the shaded region in Fig.~\ref{fig:all_self_similar_flows}. By solving for the whole flow, these two bounds translate into the two solid black lines that confine the green region in Fig.~\ref{fig:max_alpha_vs_xi_wall}. The lower (upper) bound in \eqref{bound_hibrids} corresponds to the upper (lower) bound on $\xi_w$ in Fig.~\ref{fig:max_alpha_vs_xi_wall}. Notice that the bound on deflagrations happens to precisely match the lower bound on hybrids and therefore these two types of flows do not coexist. The matching of the two bounds could be expected as deflagrations are bounded by the flows that fulfill $v_+=c_s$, which is one of the defining conditions for hybrid flows. 
Thus deflagrations and hybrids do not compete in the same region of parameters.

We see that, as we increase the strength factor, the first solutions that cease to exist are the hybrid flows. The next ones to be forbidden are detonations. Therefore, for strong transitions only deflagrations survive, but they are confined to regions with very low wall velocities.

This picture is opposite to that in the  supercooled case \citep{Espinosa:2010hh}, where deflagrations cease to exist  for strong transitions while hybrids and detonations become confined to higher and higher wall velocities.

In Fig.~\ref{fig:max_alpha_vs_xi_wall} we have added a long-dashed red curve separating the flows for which the pressure inside the bubble is higher (flows above the curve) or lower (flows below the curve) than the pressure outside the bubble. We see that, for a given wall velocity, these two cases occur for phase transitions  with high or low strength factor, respectively. 
Motivated by these findings, we have explored the possible orderings of the pressures for a supercooled phase transition, where the standard picture is that the  inside pressure is always higher than the outside pressure. We have constructed a large family of supercooled flows and we have found no counterexample to this picture. 

The bounds above are model-independent, in the sense that they follow only from the allowed regions of \fig{fig:all_self_similar_flows} and from the existence of a minimum energy density in each phase. As we now discuss, there are two additional bounds that depend on some details of the model, meaning on the numerical coefficients $a_{H,L}$, $b_{H,L}$ and $c_{H,L}$ that specify the EoS.

The first bound follows from the existence of an upper limit on $\alpha_n$ given by the minimum energy density possible for nucleation, since nucleation can only happen in the stable region of the low-energy phase. This minimum will be attained at some point on the curve of first-order phase transitions, so we have
\begin{equation}
	\alpha_n \leq \frac{\epsilon}{\mathrm{Min}(e_L(T_c,\mu_c))}.
\end{equation}
This bound would correspond to a horizontal line in Fig. \ref{fig:max_alpha_vs_xi_wall}. We do not show it in the figure since its position would depend on all the numerical coefficients in the EoS, but we note that it may compete with the other bounds presented in this section.

The second bound follows from entropic considerations. The flows presented above are described by ideal hydrodynamics, and therefore they  do not produce entropy except at the discontinuities. At each discontinuity the entropy production must be non-negative. By Stoke's theorem this amounts to the requirement that the entropy flow is non-negative across the discontinuity, which results in the condition 
\begin{equation}
 \gamma_{+} v_{+} s_{+}-\gamma_{-} v_{-} s_{-}\geq 0  \,.
\label{Entropy_across_discontinuity}
\end{equation}
The entropy densities on either side of the discontinuity, $s_{\pm}$, depend on the numerical coefficients that specify the EoS. The flows allowed by this bound for $\mu=0$ and  \mbox{$a_L/a_H=\{0.5,0.75,0.9\}$}   are those below the corresponding discontinuous curves in \fig{fig:max_alpha_vs_xi_wall}. We have verified that, for $a_L/a_H=0.5$, the entropy production is positive for the flows in 
Figs.~\ref{fig:bubbles_conformal} and \ref{fig:bubbles_conformal_p_and_e}. 
In all the cases that we have examined we have found that the most restrictive bound is the entropic bound. It would be interesting to investigate if this is true in general.

Notice that flows that produce negative entropy are not physically meaningful as expanding, superheated bubbles. However, their time reversal gives rise to physically meaningful flows with positive entropy production that can be interpreted as contracting ``drops'' that were studied in  \cite{Rezzolla:1995kv,Rezzolla:2013dea}.

\section{Efficiency factor}
Ref.~\cite{Casalderrey-Solana:2022rrn} showed that the dynamics of superheated bubbles in NS mergers can give rise to an interesting GW signal in the MHz range. A relevant quantity to estimate the amount of  GW production is the so called efficiency factor \cite{Hindmarsh:2020hop}, which is a measure of how much energy has been deposited in the motion of the fluid. The energy fraction can always be computed as the fraction of kinetic energy to the total contained in the volume of the the expanding bubble,

\begin{equation}
    \kappa = \frac{\int_{V_{\text{bubble}}}d^3x v T^{tr}}{\int_{V_{\text{bubble}}}d^3x T^{tt}} = \frac{\int_0^{\xi_{\text{max}}}d\xi \xi^2 w \gamma^2 v^2}{\int_0^{\xi_{\text{max}}}d\xi \xi^2\left(w\gamma^2-p\right)} = \frac{3\alpha_n}{\epsilon \xi_{\text{max}}^3}\int_{0}^{\xi_{\text{max}}}d\xi\xi^2w\gamma^2v.
    \label{eq:efficiency-factor}
\end{equation}
To obtain the shown expression we have used the fact that the volume of the bubble is $V_{\text{bubble}} = 4/3\pi\xi_{\text{max}}^3t^3$, with $\xi_{\text{max}}=c_s$ ($\xi_{\text{max}}=v_w$) for deflagrations and hybrids (detonations), and replaced the integral in the denominator by the total energy in the original state, before nucleation.

\begin{figure}
	\centering
\includegraphics[width=0.85\textwidth]{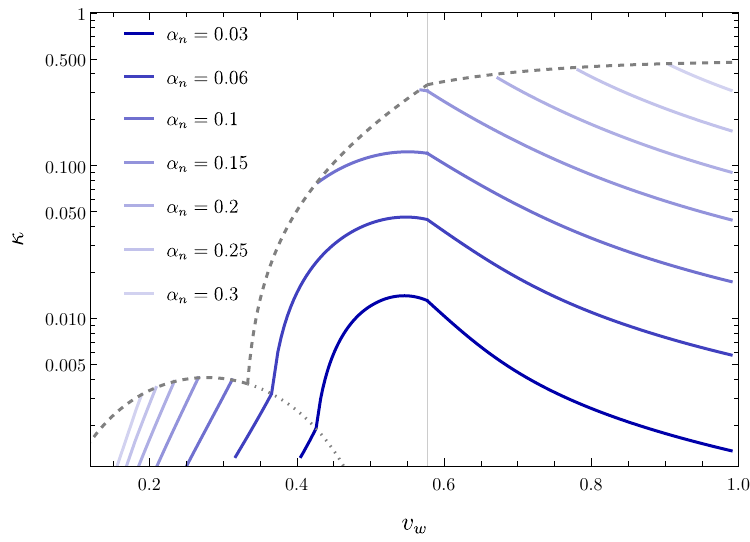}
\caption{Efficiency factor \eqref{eq:efficiency-factor} as a function of the bubble wall velocity for several choices of the strength factor. The dashed gray line represents the line beyond which no solution is allowed. The dotted gray line represents the boundary between deflagrations and hybrids, while the solid vertical line lies at the speed of sound, where hybrids transition into detonations.}
	\label{fig:kappa_vs_xi_wall}
\end{figure}

We have computed the efficiency factor for several strength factors as a function of the wall velocity in the allowed regions where we can find bubble solutions. The result is summarized in Fig.~\ref{fig:kappa_vs_xi_wall}. The empty regions above the gray dashed line arise from the disallowed regions, above the solid black curves, in \fig{fig:max_alpha_vs_xi_wall}.

The behavior of the efficiency factor for deflagrations and detonations is quite simple, since it is monotonous in both cases. For deflagrations the efficiency factor increases with the wall velocity at fixed strength, whereas for detonations it decreases. In both cases the efficiency factor increases with the strength factor at fixed wall velocity (until one hits the maximum possible $\kappa$ in dashed gray).

Hybrids exhibit a richer, non-monotonic behaviour. For low $\alpha_n$, hybrids connect the deflagrations (along the dotted gray line)  with the detonations (at the vertical $v_w = c_s$ solid line) and the efficiency factor has a maximum at some intermediate value of the wall velocity. As the envelope -- in dashed gray -- shows, the largest efficiency factor value is found for detonations, increasing with the wall speed up to a maximum value slightly below $0.5$.

Fig.~\ref{fig:kappa_vs_xi_wall} includes all possible flows allowed by  hydrodynamics. However, it must be kept in mind that there are more restrictive, model-dependent bounds that will  exclude some of these flows, which in turn may reduce the possible maximum value of the efficiency factor.

\section{Large jump in the number of degrees of freedom}

We now turn our attention to the particular case in which the jump in the number of degrees of freedom between the two phases is large, that is, $e_L \ll e_H$. In the supercooled case, this analysis led to an estimate for the wall velocity as a function of the nucleation temperature \cite{Sanchez-Garitaonandia:2023zqz}. We will obtain an analogous result in the superheated case. 

A convenient way to parameterize a large jump in the number of degrees of freedom is to rescale the $L$-phase pressure as $p_L \to x \, p_L$, with $x\ll 1$. This immediately implies that all extensive quantities in the $L$-phase scale as $x$. In turn, this means that the strength factor diverges as 
\begin{equation}
\alpha_n = \frac{\epsilon}{e_L} \sim x^{-1} \,.
\end{equation}
In view of \fig{fig:max_alpha_vs_xi_wall}, this immediately tells us that the flow must be a deflagration with very small wall velocity, since this is the only type of flow that exists for arbitrarily large strength factors. We have verified that the black solid curve on the left of  \fig{fig:max_alpha_vs_xi_wall}  at small $\xi_w$ is given by 
\begin{equation}
\label{tango}
\xi_w \sim \alpha_n^{-1} \,.
\end{equation}
Comparing the two equations above we conclude that the velocity must scale at small $x$ as
\begin{equation}
\label{last}
\xi_w \sim x \,.
\end{equation}

This can also be seen from the following argument. In the limit $x\to 0$ the critical temperature and chemical potential approach finite, $x$-independent values $T_c, \mu_c$. At temperatures and chemical potentials that are not parametrically separated from these  the ratios of the following quantities in front and behind the wall obey
\begin{equation}
	\frac{e_+}{e_-} \sim \frac{s_+}{s_-} \sim \frac{w_+}{w_-} 
 \sim x \,.
\end{equation}
The pressure jump works slightly differently. By definition, $p_H(T_c,\mu_c)=p_L(T_c,\mu_c)$. This means that near the critical curve not only $p_L$ but also $p_H$ are $\mathcal{O}(x)$, and therefore so is their difference. It then follows that 

\begin{equation}
	v_+v_- = \frac{p_+-p_-}{e_-} \sim x \,,\quad 
 \frac{v_+}{v_-} = \frac{e_-}{e_++p_-}  \sim x^{-1} \,.
\end{equation}
From this we conclude that fluid velocity behind the wall in the rest frame of the wall is  $v_- \sim x$. This  is related to the  velocity in the lab frame, $v$, through  
\begin{equation}
v_- = \frac{v-\xi_w}{1-v\xi_w} \,.
\end{equation}
For deflagrations $v=0$ behind the wall and hence 
\begin{equation}
\xi_w \sim \vert v_- \vert  \sim x \,,
\end{equation}
in agreement with \eqq{last}.

\section{Discussion}
\label{disc}
Motivated by their possible role in NS mergers, in this paper we have studied the hydrodynamics of relativistic superheated bubbles in the presence of a conserved charge akin to baryon number. Equations (\ref{eq:Self_Similar_Flow}) determine the possible flow profiles. In general, these equations are coupled to one another and one must  solve for the velocity and the charge density simultaneously. Here we have observed that this problem simplifies if the speed of sound is constant in each branch of the EoS. In this case one can solve sequentially for $v, w$ and $n$, meaning that the velocity profile is unaffected by the presence of the charge. For simplicity, in this paper we have restricted ourselves to this  case by choosing a bag-model EoS for which $c_s^2=1/3$ in both branches. One tantalizing result is that, in the case of a large jump in the number of degrees of freedom between the two phases,  only deflagrations are possible and the bubble wall velocity is related to the strength of the transition through \eqn{tango}. Extrapolating to the hadronic-quark matter transition in QCD this would predict a value for the wall velocity around $\xi_w\sim 0.1$, in agreement with the  ballpark value suggested in \cite{Casalderrey-Solana:2022rrn}. It would be  
interesting to verify these estimates with a more realistic approximation to the QCD EoS. In this case the velocity and the charge profiles will not decouple from one another and the fluid flows could be qualitatively different. 

We have found two qualitative differences between superheated and  supercooled flows. The first one is that some superheated detonations and hybrids can develop a necessarily metastable region behind the bubble wall, which at sufficiently long times will decay by nucleating bubbles of the $L$-phase. 
The second difference is that the pressure inside an expanding superheated bubble (at $\xi=0$) can be higher or lower than the pressure asymptotically far away from the bubble (at $\xi=1$). Even in the latter case, the bubble still expands by accreting energy from the outside. This raises an interesting dynamical question about how this steady state is reached from the slightly over-critical bubble that is initially nucleated. The reason is that, for the  critical bubble, the inside pressure must be slightly higher than the outside pressure, since it must exactly compensate for the effect of the latter plus the surface tension. It would be interesting to see the time evolution of the inside pressure along the first instants after nucleation in a microscopic model, following the analysis in the supercooled case \cite{Bea:2021zsu}. An additional insight provided by such a model would be a prediction of the bubble wall velocity from the  microscopic dynamics. We will report on these aspects elsewhere.

\section*{Acknowledgements}
We thank Alexandre Serantes for useful discussions. 
This work was supported by the “Center of Excellence Maria de Maeztu 2020-2023” award to the ICCUB (CEX2019-000918-M) funded by MCIN/AEI/10.13039/501100011033. YB, DM and JCS acknowledge support from grants PID2019- 105614GB-C21, PID2019-105614GB-C22, PID2022-136224NB-C21, PID2022-136224NB-C22, and 2021-SGR-872. The work of YB is also funded by a  ``Beatriu de Pin\'os'' postdoctoral program under the Ministry of Research and Universities of the Government of Catalonia (2022 BP 00225) and by a ``Maria Zambrano'' postdoctoral fellowship from the University of Barcelona.  The work of MSG is supported by the European Research Council (ERC) under the European Union's Horizon 2020 research and innovation program (grant agreement No758759).

\appendix
\section{Non-relativistic, superheated bubbles with an incompressible phase}
\label{AppA}
Superheated bubbles occur in many everyday situations, such as the formation of vapor bubbles in superheated water. In these cases, the matter surrounding the bubble walls is non-relativistic. Additionally, for the relevant range of pressures, the matter outside the superheated bubbles behaves as an incompressible fluid. These two features make the dynamics of those superheated bubbles very different from their relativistic analogues, as described in this work. Specifically, in the non-relativistic case, the bubble size does not grow at a constant rate; instead, the growth rate decreases with time $t^{-1/2}$. For completeness, this appendix reviews the dynamics of non-relativistic superheated bubbles, and the discussion here follows closely the review \cite{vbblreview}

We assume a non-relativistic system in its low-temperature phase that has been superheated past its critical temperature. In this system, characterized by a mass density $\rho_L$, the low-temperature phase is incompressible, meaning $\rho_L$ remains constant. When a bubble of the thermodynamically stable high-temperature phase nucleates, it grows due to entropic preference. This phase has its own mass density, $\rho_H$. However, similar to vapor bubbles in superheated water, we do not assume the fluid inside the bubble is incompressible. As with the relativistic case, once the bubble grows sufficiently large, the dynamics inside and outside the bubble can be described by non-relativistic hydrodynamics. The interface between the two phases can then be approximated as a discontinuity in the hydrodynamic fields.

Paralleling the discussion of the relativistic case, we begin by characterizing the fluid velocity in the vicinity of the bubble. This is analogous to the velocity profiles presented in fig.~(\ref{fig:all_self_similar_flows}). In the low-temperature phase, the mass continuity equation combined with the incompressibility condition implies that the velocity field is divergenceless, i.e., ${\bf \nabla v}=0$.
Under spherical symmetry, the velocity field outside the bubble is radial and can be expressed as:
\begin{equation}
{\bf v} (r,t)=\frac{F(t)}{r^2} \hat r \, \quad \quad r\geq R(t)
\label{eq:vfgen}
\end{equation}
where $R(t)$ is the bubble radius and $F(t)$ is a function determined by mass conservation, analogous to the charge conservation equation in non-relativistic systems. The velocity field vanishes far from the bubble, ensuring that the fluid at a distance remains at rest.

Since the bubble wall does not accumulate mass, the mass flux on both sides of the wall must be equal. This is equivalent to the last matching condition in Eq. \eqref{eq:matchingc}.Imposing that the matter inside the bubble is at rest, and considering that as the bubble grows, the change in density between the high and low-temperature phases must be compensated by the influx of mass from the low-temperature phase outside the wall:
\begin{equation}
\left(\rho_L - \rho_H\right) \dot R(t)=\rho_L v (R(t),t) \Rightarrow F(t)= R^2(t) \dot R(t) \left(1-\frac{\rho_H}{\rho_L}\right)
\label{eq:Fform}
\end{equation}
This expression for the velocity field, combined with the Navier-Stokes (NS) equation for incompressible fluids,
\begin{equation}
-{\bf \nabla} \left(\frac{p}{\rho_L}\right)= \frac{\partial \bf{ v}}{\partial t} + \left({\bf v} \cdot {\bf \nabla} \right) {\bf v} - \nu \nabla^2 {\bf v}
 \end{equation}
where $\nu=\eta/\rho_L$ is the kinematic viscosity, determines the form of the pressure field in front of the bubble. By substituting Eqs. \eqref{eq:vfgen}
 and \eqref{eq:Fform} into the NS equation and integrating in the radial direction, we obtain the pressure difference in the low-temperature phase both far away from the bubble, $p_\infty$ and next to the bubble wall $p_L(R)$:
\begin{equation}
\label{eq:almostRPeq}
\frac{p_L(R)-p_\infty}{\rho_F}=(1-Q) \left(\frac{1}{2} (3+Q) \dot R(t)^2+R(t) \ddot R(t)\right)
\end{equation}
with $Q=\equiv\rho_H/\rho_L$.
The pressure outside the bubble can be related to the pressure inside the bubble through the non-relativistic analog of the matching conditions for the momentum flux across the wall, given by the first equation in Eq. \eqref{eq:matchingc}. Taking into account the effect of the bubble's surface tension, $\gamma$, the fluid stress on both sides of the wall is given by:
\begin{equation}
p_L(R)=p_H - 2\frac{\gamma}{R} -4 (1-Q)\frac{\nu}{R} \dot R
\end{equation}
where the last term comes from the sheer stress contribution of the velocity field, Eq. \eqref{eq:Fform}. 
For very large bubbles, the effect of the surface tension and the viscosity becomes negligible; this is the reason why we have not considered those contribution in the relativistic analysis. Nevertheless, keeping those terms, and together with 
 \eqq{eq:almostRPeq}, we obtain the so called Rayleigh-Plesset (RP) equation
\begin{equation}
\label{eq:deltaPRP}
\frac{p_H-p_\infty}{\rho_L}=(1-Q) \left(\frac{1}{2} (3+Q) \dot R(t)^2+R(t) \ddot R(t) +4 \frac{\nu}{R} \dot R\right) + 2\frac{\gamma}{R} 
\end{equation}
which governs the dynamics of non-relativistic bubbles in incompressible fluids.

To complete the analysis, we also must consider the effect of the energy flux across the bubble wall, which in the relativistic case is given by the middle equation in Eq.~\eqref{eq:matchingc}. In general, in the non-relativistic limit, the energy flux  is given by 
\begin{equation}
\label{eq:EFIF}
{\bf \mathcal{F}}= \rho \left(\frac{1}{2} v^2 + {\bar w}\right) {\bf v} - \eta {\bf v} \cdot \sigma - \kappa {\bf \nabla} T
\end{equation}
with $\bar w=\bar e +\frac{p}{\rho}$  and 
${\bar e}$ are the enthalpy and energy per unit mass, $\sigma$ is the sheer tensor and $\kappa$ the heat conductivity. For spherical bubbles and using Eqs.~\eqref{eq:vfgen} and~\eqref{eq:Fform}, the radial component of the flux right outside of the bubble wall is given by
\begin{equation}
\label{eq:hf}
\mathcal{F}_{\rm outside}^r=  \frac{1}{2}(1-Q)^3 \rho \dot R^3 + \rho \bar w (1-Q) \dot R - 4 \eta (1-Q)^2 \frac{\dot R^2}{R} - \kappa \frac{\partial  T}{ \partial r}  
\end{equation}
This expression shows that non-relativistic bubbles cannot grow at a constant rate, as their relativistic analogues do, at least for normal fluids, such that the density of the fluid phase is larger than the density of the vapour phase, $Q<1$. For a constant and positive $\dot R$, the first to terms of the expression above dominate and the energy flux is positive, which implies that the fluid motion outside the wall takes away energy from the interior of the bubble. This can also inferred from the combination of Eqs.~\eqref{eq:vfgen} and~\eqref{eq:Fform}, which show that the velocity of the fluid outside of the bubble is positive, unlike the relativistic case. Therefore, the only way that the latent heat can be compensated is by reducing the internal energy 
 of the fluid outside; when this energy arrives to the critical energy, the pressure difference in the inside and outside of the bubble become equal and the fluid velocity must vanish, as inferred from \eqq{eq:deltaPRP}. Therefore, superheated bubbles cannot expand at constant velocities for arbitrarily long times. 

 For bubbles to continue growing at asymptotic late times, energy must be transferred toward the interior of the bubble. This is achieved by heat transfer, represented by the last term in Eq.~\eqref{eq:hf}, which becomes the dominant mechanism at these late times. In this regime, integrating over the bubble's surface, the energy balance between the heat flux and the latent heat can be expressed as follows:
 \begin{equation}
 L\frac{d}{dt} \left(\frac{4}{3} \pi R^2 \rho_H\right) =4 \pi R^3 \kappa \left.  \frac{\partial T}{\partial r} \right|_{r=R(t)}
 \end{equation}
with $L$ the latent heat of the transition. In this limit, the temperature gradient outside the bubble is primarily governed by heat diffusion, yielding $\frac{\partial T}{\partial r}\sim \Delta T/(D t)^{1/2}$ with $\Delta T$, where $\Delta T$ is the temperature difference between the wall and the ambient, and $D$ is the thermal diffusivity of the fluid outside the wall. Consequently, in this regime, the bubble radius $R$ scales as $R\sim t^{1/2}$. For a comprehensive derivation of this expression, refer to \cite{vbblreview} and the references therein.

\newpage
\bibliographystyle{utphys}
\bibliography{refsHolographicBubbles.bib}

\providecommand{\href}[2]{#2}\begingroup\raggedright\begin{thebibliography}{10}

\bibitem{Stephanov:2004wx}
M.~A. Stephanov, ``{QCD Phase Diagram and the Critical Point},''
  \href{http://dx.doi.org/10.1143/PTPS.153.139}{{\em Prog. Theor. Phys. Suppl.}
  {\bfseries 153} (2004) 139--156},
  \href{http://arxiv.org/abs/hep-ph/0402115}{{\ttfamily arXiv:hep-ph/0402115}}.

\bibitem{Alford:2007xm}
M.~G. Alford, A.~Schmitt, K.~Rajagopal, and T.~Sch\"afer, ``{Color
  superconductivity in dense quark matter},''
  \href{http://dx.doi.org/10.1103/RevModPhys.80.1455}{{\em Rev. Mod. Phys.}
  {\bfseries 80} (2008) 1455--1515},
  \href{http://arxiv.org/abs/0709.4635}{{\ttfamily arXiv:0709.4635 [hep-ph]}}.

\bibitem{Fukushima:2010bq}
K.~Fukushima and T.~Hatsuda, ``{The phase diagram of dense QCD},''
  \href{http://dx.doi.org/10.1088/0034-4885/74/1/014001}{{\em Rept. Prog.
  Phys.} {\bfseries 74} (2011) 014001},
  \href{http://arxiv.org/abs/1005.4814}{{\ttfamily arXiv:1005.4814 [hep-ph]}}.

\bibitem{Guenther:2022wcr}
J.~N. Guenther, ``{An overview of the QCD phase diagram at finite $T$ and
  $\mu$},'' in {\em {38th International Symposium on Lattice Field Theory}}.
\newblock 1, 2022.
\newblock \href{http://arxiv.org/abs/2201.02072}{{\ttfamily arXiv:2201.02072
  [hep-lat]}}.

\bibitem{Casalderrey-Solana:2022rrn}
J.~Casalderrey-Solana, D.~Mateos, and M.~Sanchez-Garitaonandia, ``{Mega-Hertz
  Gravitational Waves from Neutron Star Mergers},''
  \href{http://arxiv.org/abs/2210.03171}{{\ttfamily arXiv:2210.03171
  [hep-th]}}.

\bibitem{Most:2018eaw}
E.~R. Most, L.~J. Papenfort, V.~Dexheimer, M.~Hanauske, S.~Schramm,
  H.~St\"ocker, and L.~Rezzolla, ``{Signatures of quark-hadron phase
  transitions in general-relativistic neutron-star mergers},''
  \href{http://dx.doi.org/10.1103/PhysRevLett.122.061101}{{\em Phys. Rev.
  Lett.} {\bfseries 122} no.~6, (2019) 061101},
  \href{http://arxiv.org/abs/1807.03684}{{\ttfamily arXiv:1807.03684
  [astro-ph.HE]}}.

\bibitem{Most:2019onn}
E.~R. Most, L.~Jens~Papenfort, V.~Dexheimer, M.~Hanauske, H.~Stoecker, and
  L.~Rezzolla, ``{On the deconfinement phase transition in neutron-star
  mergers},'' \href{http://dx.doi.org/10.1140/epja/s10050-020-00073-4}{{\em
  Eur. Phys. J. A} {\bfseries 56} no.~2, (2020) 59},
  \href{http://arxiv.org/abs/1910.13893}{{\ttfamily arXiv:1910.13893
  [astro-ph.HE]}}.

\bibitem{Ecker:2019xrw}
C.~Ecker, M.~J\"arvinen, G.~Nijs, and W.~van~der Schee, ``{Gravitational waves
  from holographic neutron star mergers},''
  \href{http://dx.doi.org/10.1103/PhysRevD.101.103006}{{\em Phys. Rev. D}
  {\bfseries 101} no.~10, (2020) 103006},
  \href{http://arxiv.org/abs/1908.03213}{{\ttfamily arXiv:1908.03213
  [astro-ph.HE]}}.

\bibitem{Prakash:2021wpz}
A.~Prakash, D.~Radice, D.~Logoteta, A.~Perego, V.~Nedora, I.~Bombaci,
  R.~Kashyap, S.~Bernuzzi, and A.~Endrizzi, ``{Signatures of deconfined quark
  phases in binary neutron star mergers},''
  \href{http://dx.doi.org/10.1103/PhysRevD.104.083029}{{\em Phys. Rev. D}
  {\bfseries 104} no.~8, (2021) 083029},
  \href{http://arxiv.org/abs/2106.07885}{{\ttfamily arXiv:2106.07885
  [astro-ph.HE]}}.

\bibitem{Weih:2019xvw}
L.~R. Weih, M.~Hanauske, and L.~Rezzolla, ``{Postmerger Gravitational-Wave
  Signatures of Phase Transitions in Binary Mergers},''
  \href{http://dx.doi.org/10.1103/PhysRevLett.124.171103}{{\em Phys. Rev.
  Lett.} {\bfseries 124} no.~17, (2020) 171103},
  \href{http://arxiv.org/abs/1912.09340}{{\ttfamily arXiv:1912.09340 [gr-qc]}}.

\bibitem{Tootle:2022pvd}
S.~Tootle, C.~Ecker, K.~Topolski, T.~Demircik, M.~J\"arvinen, and L.~Rezzolla,
  ``{Quark formation and phenomenology in binary neutron-star mergers using
  V-QCD},'' \href{http://arxiv.org/abs/2205.05691}{{\ttfamily arXiv:2205.05691
  [astro-ph.HE]}}.

\bibitem{Hindmarsh:2020hop}
M.~B. Hindmarsh, M.~L\"uben, J.~Lumma, and M.~Pauly, ``{Phase transitions in
  the early universe},''
  \href{http://dx.doi.org/10.21468/SciPostPhysLectNotes.24}{{\em SciPost Phys.
  Lect. Notes} {\bfseries 24} (2021) 1},
  \href{http://arxiv.org/abs/2008.09136}{{\ttfamily arXiv:2008.09136
  [astro-ph.CO]}}.

\bibitem{Barni:2024lkj}
G.~Barni, S.~Blasi, and M.~Vanvlasselaer, ``{The hydrodynamics of inverse phase
  transitions},'' \href{http://arxiv.org/abs/2406.01596}{{\ttfamily
  arXiv:2406.01596 [hep-ph]}}.

\bibitem{Espinosa:2010hh}
J.~R. Espinosa, T.~Konstandin, J.~M. No, and G.~Servant, ``{Energy Budget of
  Cosmological First-order Phase Transitions},''
  \href{http://dx.doi.org/10.1088/1475-7516/2010/06/028}{{\em JCAP} {\bfseries
  06} (2010) 028}, \href{http://arxiv.org/abs/1004.4187}{{\ttfamily
  arXiv:1004.4187 [hep-ph]}}.

\bibitem{Komoltsev:2021jzg}
O.~Komoltsev and A.~Kurkela, ``{How Perturbative QCD Constrains the Equation of
  State at Neutron-Star Densities},''
  \href{http://dx.doi.org/10.1103/PhysRevLett.128.202701}{{\em Phys. Rev.
  Lett.} {\bfseries 128} no.~20, (2022) 202701},
  \href{http://arxiv.org/abs/2111.05350}{{\ttfamily arXiv:2111.05350
  [nucl-th]}}.

\bibitem{Caprini:2011uz}
C.~Caprini and J.~M. No, ``{Supersonic Electroweak Baryogenesis: Achieving
  Baryogenesis for Fast Bubble Walls},''
  \href{http://dx.doi.org/10.1088/1475-7516/2012/01/031}{{\em JCAP} {\bfseries
  01} (2012) 031}, \href{http://arxiv.org/abs/1111.1726}{{\ttfamily
  arXiv:1111.1726 [hep-ph]}}.

\bibitem{Ares:2020lbt}
F.~R. Ares, M.~Hindmarsh, C.~Hoyos, and N.~Jokela, ``{Gravitational waves from
  a holographic phase transition},''
  \href{http://dx.doi.org/10.1007/JHEP04(2021)100}{{\em JHEP} {\bfseries 21}
  (2020) 100}, \href{http://arxiv.org/abs/2011.12878}{{\ttfamily
  arXiv:2011.12878 [hep-th]}}.

\bibitem{Rezzolla:1995kv}
L.~Rezzolla, J.~C. Miller, and O.~Pantano, ``{Evaporation of quark drops during
  the cosmological quark - hadron transition},''
  \href{http://dx.doi.org/10.1103/PhysRevD.52.3202}{{\em Phys. Rev. D}
  {\bfseries 52} (1995) 3202--3213},
  \href{http://arxiv.org/abs/astro-ph/9502064}{{\ttfamily
  arXiv:astro-ph/9502064}}.

\bibitem{Rezzolla:2013dea}
L.~Rezzolla and O.~Zanotti,
  \href{http://dx.doi.org/10.1093/acprof:oso/9780198528906.001.0001}{{\em
  {Relativistic Hydrodynamics}}}.
\newblock Oxford University Press, 9, 2013.

\bibitem{Sanchez-Garitaonandia:2023zqz}
M.~Sanchez-Garitaonandia and J.~van~de Vis, ``{Prediction of the bubble wall
  velocity for a large jump in degrees of freedom},''
  \href{http://arxiv.org/abs/2312.09964}{{\ttfamily arXiv:2312.09964
  [hep-ph]}}.

\bibitem{Bea:2021zsu}
Y.~Bea, J.~Casalderrey-Solana, T.~Giannakopoulos, D.~Mateos,
  M.~Sanchez-Garitaonandia, and M.~Zilh\~ao, ``{Bubble wall velocity from
  holography},'' \href{http://dx.doi.org/10.1103/PhysRevD.104.L121903}{{\em
  Phys. Rev. D} {\bfseries 104} no.~12, (2021) L121903},
  \href{http://arxiv.org/abs/2104.05708}{{\ttfamily arXiv:2104.05708
  [hep-th]}}.

\bibitem{vbblreview}
A.~Prosperetti, ``Vapor bubbles,''
  \href{http://dx.doi.org/https://doi.org/10.1146/annurev-fluid-010816-060221}{{\em
  Annual Review of Fluid Mechanics} {\bfseries 49} no.~Volume 49, 2017, (2017)
  221--248}.
  \url{https://www.annualreviews.org/content/journals/10.1146/annurev-fluid-010816-060221}.

\end{thebibliography}\endgroup

\end{document}